\documentclass[useAMS,usenatbib,fleqn]{mn2e}
\usepackage{listings}
\usepackage{amsmath}
\usepackage{natbib}
\usepackage{graphicx}
\usepackage{color}
\usepackage{rotating}
\usepackage{nicefrac}
\usepackage{txfonts}
\usepackage{subfigure}
\usepackage{trivfloat}
\trivfloat{listfloat}
\usepackage{longtable}
\usepackage[percent]{overpic}
\usepackage{xspace}
\usepackage{url}
\usepackage{mathrsfs}
\usepackage{amssymb}
\usepackage{mathtools} 
\usepackage{enumitem}
\usepackage[percent]{overpic}

\usepackage{epsfig}
\usepackage{amsmath}
\usepackage{rotating}
\usepackage{natbib}
\usepackage{multirow}
\usepackage[noend]{algpseudocode} 

\newcommand\eps{\ensuremath{\mathit{eps}}}

\DeclareMathOperator{\Ha}{\mathcal{H}}
\newcommand{\whf}{{\sc \tt WHFast}\xspace}

\lstdefinestyle{customc}{
  belowcaptionskip=1\baselineskip,
  breaklines=true,
  language=C,
  showstringspaces=false,
  basicstyle=\footnotesize\ttfamily,
}

\usepackage{array}
\usepackage{appendix}
\usepackage{comment}

\def \aap{A\&A}

\def \apjl{ApJ}
\def \apjs{ApJS}

\newcommand{\reb}{{\sc \tt REBOUND}\xspace}
\newcommand{\mer}{{\sc \tt MERCURY}\xspace}
\newcommand{\swifter}{{\sc \tt SWIFTER}\xspace}
\newcommand{\wh}{{\sc \tt WH}\xspace}
\newcommand{\swifterwhm}{{\sc \tt SWIFTER-WHM}\xspace}
\newcommand{\swifterhelio}{{\sc \tt SWIFTER-HELIO}\xspace}
\newcommand{\swiftertu}{{\sc \tt SWIFTER-TU4}\xspace}

\def\gsim{\;\rlap{\lower 2.5pt
 \hbox{$\sim$}}\raise 1.5pt\hbox{$>$}\;}
\def\lsim{\;\rlap{\lower 2.5pt
   \hbox{$\sim$}}\raise 1.5pt\hbox{$<$}\;}

\title[\whf: A fast and unbiased Wisdom-Holman integrator]{\whf: A fast and unbiased implementation of a symplectic Wisdom-Holman integrator for long term gravitational simulations}
\date{Submitted: 8th May 2015. Accepted: 2nd June 2015.}

\author[Hanno Rein and Daniel Tamayo]{
Hanno Rein$^{1,2}$ and Daniel Tamayo$^{1,3,4}$\\ 
$^1$ Department of Physical and Environmental Sciences, University of Toronto at Scarborough, Toronto, Ontario M1C 1A4, Canada\\
$^2$ Department of Astronomy and Astrophysics, University of Toronto, Toronto, Ontario, M5S 3H4, Canada\\
$^3$ Canadian Institute for Theoretical Astrophysics, 60 St. George St, University of Toronto, Toronto, Ontario M5S 3H8, Canada\\
$^4$ Centre for Planetary Sciences Fellow
}

\vspace{0.5\baselineskip}

\begin{document}
\maketitle

\begin{abstract}
We present \whf, a fast and accurate implementation of a Wisdom-Holman symplectic integrator for long-term orbit integrations of planetary systems.
\whf is significantly faster and conserves energy better than all other Wisdom-Holman integrators tested.
 
We achieve this by significantly improving the Kepler-solver and ensuring numerical stability of coordinate transformations to and from Jacobi coordinates.
These refinements allow us to remove the linear secular trend in the energy error that is present in other implementations.
For small enough timesteps we achieve Brouwer's law, i.e. the energy error is dominated by an unbiased random walk due to floating-point round-off errors.

We implement symplectic correctors up to order eleven that significantly reduce the energy error.
We also implement a symplectic tangent map for the variational equations.
This allows us to efficiently calculate two widely used chaos indicators the Lyapunov characteristic number (LCN) and the Mean Exponential Growth factor of Nearby Orbits~(MEGNO).  

\whf is freely available as a flexible C package, as a shared library, and as an easy-to-use python module.

\end{abstract}


\section{Introduction}
\label{sec:intro}
Celestial mechanics, the field that deals with the motion of celestial objects, has been an active field of research since the days of Newton and Kepler.
Analytic solutions only exist for a few special cases.
Historically, the main driver for the development of perturbation theory has been the problem of planets orbiting the Sun.  
Because the central body is so much more massive than the planets, it is profitable to ask how the small mutual tugs between the planets modify the Keplerian orbits they would each individually follow around the Sun in the absence of the other bodies.  
This analytical approach has been, and continues to be, successful in explaining many important features of planetary orbits.
However, the Solar System is chaotic, and the rise of computing power has yielded many important insights.
There is therefore considerable interest in developing fast and accurate numerical integrators.  

A large number of such integrators have been developed over the years to perform this task.
For many long term integrations, symplectic integrators have proven to be a favourable choice.
Symplectic schemes incorporate the symmetries of Hamiltonian systems, and therefore typically conserve quantities like the energy and angular momentum better than non-symplectic integrators.

For integrations of planetary systems, \cite{WisdomHolman1991}, and independently \cite{Kinoshita1991}, developed a widely used class of symplectic integrators. 
The ideas of \cite{WisdomHolman1991} developed from the original ideas of the mapping method of \cite{Wisdom1981}.
We refer to these as a Wisdom-Holman mapping or a Wisdom-Holman integrator.
Since then, many authors have modified and built upon this method, and several have made their integrators publicly available to the astrophysics community \citep[e.g.][]{Chambers1997,Duncan1998}.

The Wisdom-Holman integrator exploits the intuition from perturbation theory that one can separate the problem into a system of Keplerian orbits about the Sun, modified by small perturbations among the planets. 
The nuisance is that while Newton provided us the solution to the two-body problem, Poincar{\'e} showed that the remaining superimposed perturbations are not integrable.
Analytically, the traditional way forward is to average over the short-period oscillations in the problem to yield approximate solutions.  
The great insight of Wisdom and Holman was that, at the same level of approximation, one can {\it add} high frequency terms.
By judicious choice of these additional frequencies, the perturbations among the planets can be transformed into trivially integrated delta functions.
The result is an exceedingly efficient integrator that has proven an indispensable tool for modern studies in celestial mechanics.

In this paper, we present results from a complete reimplementation of the Wisdom-Holman integrator.
We show how to speed up the algorithm in several ways and dramatically increase its accuracy.
Many of the improvements are related to finite double floating-point precision on modern computers \citep[IEEE754,][]{IEEE754}.
The fact that almost all real numbers cannot be represented exactly in floating-point precision leads to important consequences for the numerical stability of any algorithm and the growth of numerical round-off error.

To our knowledge, we present the first publicly available Wisdom-Holman integrator that is unbiased, i.e. the errors are random and uncorrelated.
This leads to a very slow error growth.
For sufficiently small timesteps, we achieve Brouwer's law, i.e., the energy error grows as time to the power of one half.

We have also sped up the integrator through various improvements to the integrator's Kepler-solver.
Our implementation allows for the evolution of variational equations (to determine whether orbits are chaotic) at almost no additional cost.
Additionally, we implement so-called symplectic correctors up to order eleven to increase the accurary~\citep{Wisdom1996}, allow for arbitrary unit choices, and do not tie the integration to a particular frame of reference.

We make our integrator, which we call \whf, publicly available in its native C99 implementation and as an easy-to-use python module.

The remainder of this paper is structured as follows.
We first summarize the concepts and algorithms used in this paper, including Jacobi Coordinates, our choice of Hamiltonian splitting, the symplectic Wisdom-Holman map, symplectic correctors and the variational equations in Sect.~\ref{sec:background}.
We then go into detail discussing the improvements we have made to these algorithms in Sect.~\ref{sec:improvements}.
Numerical tests are presented in Sect.~\ref{sec:numericalresults} before we conclude in Sect.~\ref{sec:conclusions}.

\section{Background}\label{sec:background}
The Hamiltonian $\mathcal{H}$ of the gravitational $N$-body system can be written as the sum of kinetic and potential terms in Cartesian coordinates
\begin{eqnarray}
\mathcal{H} &=& \sum_{i=0}^{N-1} \frac{\mathbf{p}_i^2}{2m_i} - \sum_{i=0}^{N-1} \sum_{j=i+1}^{N-1} \frac{Gm_im_j}{|\mathbf{r}_i-\mathbf{r}_j|}.\label{eq:H}
\end{eqnarray}
One way forward toward separating out the two-body Keplerian Hamiltonians is to transform to heliocentric coordinates involving the centre-of-mass and the $\mathbf{r}_i-\mathbf{r}_0$.  
However, rewriting the Cartesian momenta in terms of heliocentric momenta (which have an additional component along the centre-of-mass momentum), leads to several cross-terms.  
Alternatively, Jacobi worked out a coordinate system in which the kinetic terms are particularly clean, and the kinetic energy remains a sum of squares.  
For readers that may not be familiar, and because our improved accuracy is largely due to modifications of the manner in which we transform between Cartesian and Jacobi coordinates, we briefly review them \citep[see also][]{Plummer1918,SussmanWisdom2001,solarsystemdynamics}.

%%%%%%%%%%%%%%%%%%%%%%%%%%%%%%%%%%%%%%%%%%
\subsection{Jacobi Coordinates}
Rather than reference planet positions to the central star, a planet's Jacobi coordinates are measured relative to the centre-of-mass of all bodies with lower indices.   
For concreteness, consider a system of $N$ particles with masses $m_i$, $i=0,\ldots, N-1$. 
Let $\mathbf{r}_i$ be the position vector of the $i$-th particle with respect to an arbitrary origin that is fixed in an inertial frame.
Here we assume that the particles are ordered such that $i=0$ corresponds to the central object, $i=1$ to the innermost object orbiting the central object and so on. 
The existence of such an ordering does not restrict the architecture of the system.
For example, the coordinates of an equal-mass binary with a circumbinary particle can be expressed in Jacobi coordinates. 
But note that the ordering might in general be non-unique and that it can change during an integration.
This can have important implications for a numerical scheme using Jacobi coordinates.

The Jacobi coordinate $\mathbf{r}'_i$ of the $i$-th particles is the position relative to $\mathbf{R}_{i-1}$, the centre-of-mass of all the particles interior to the $i$-th particle:
\begin{eqnarray}
\mathbf{r}'_i &=& \mathbf{r}_i - \mathbf{R}_{i-1},\quad\quad\quad\quad\text{for } \;i=1,\ldots,N-1\\
\text{where} \quad \mathbf{R}_{i} &=& \frac{1}{M_i}\sum_{j=0}^i m_j \mathbf{r}_j \quad \text{and } \quad M_{i} \;=\; \sum_{j=0}^i m_j.
\end{eqnarray}
Other quantities such as the velocity and acceleration (also the coordinates in the variational equations, see below) transform in the same way.
This is because the Jacobi coordinates are a linear function of the Cartesian coordinates, and the velocity is the time derivative of the position in both coordinates systems. 

The momenta, however, transform differently\footnote{But note that we do not need to calculate the momenta explicitly in our algorithm.}.
The momentum conjugate to $\mathbf{r}'_i$ and the corresponding Jacobi mass are given by
\begin{eqnarray}
\mathbf{p}'_i = m'_i \dot{\mathbf{r}'}_i  = m'_i \mathbf{v'}_i \quad\quad\text{   and   } \quad\quad m'_i = m_i \frac{M_{i-1}}{M_{i}}=  \frac{m_i\,M_{i-1}}{m_i + M_{i-1}}.\label{eq:p}
\end{eqnarray}
Note that the Jacobi mass $m'_i$ is the reduced mass of $m_i$ and $M_{i-1}$.
Explicit expressions for the momenta can be found by evaluating the time derivative r{Eq.~\ref{eq:p}.}

The Jacobi coordinates above are relative coordinates for $i=1,\ldots, N-1$.
For the $0$-th coordinate, a different convention is used,
\begin{eqnarray}
\mathbf{r}'_0 = \mathbf{R}_{N-1},\quad\quad\quad m'_0 = M_{N-1},\quad\quad\quad \mathbf{p}'_0 = \sum_{j=0}^{N-1} \mathbf{p}_j.
\end{eqnarray}
Thus, $\mathbf{r}'_0$ points towards the centre-of-mass of the entire system,~$\mathbf{p}'_0$~is the total momentum and $m'_0$ is the total mass.

%%%%%%%%%%%%%%%%%%%%%%%%%%%%%%%%%%%%%%%%%%
\subsection{Hamiltonian Splitting}\label{sec:splitting}
After some algebra, we can rewrite the Hamiltonian in Eq.~\ref{eq:H} in terms of the conjugate momenta of the Jacobi coordinates \citep[e.g.][]{solarsystemdynamics,SussmanWisdom2001}. 
We only rewrite the kinetic term and keep the potential term expressed as a function of the Cartesian coordinates:
\begin{eqnarray}
\mathcal{H} &=& \sum_{i=0}^{N-1} \frac{{\mathbf{p}'_i}^2}{2m'_i} - \sum_{i=0}^{N-1} \sum_{j=i+1}^{N-1} \frac{Gm_im_j}{|\mathbf{r}_i-\mathbf{r}_j|}.
\end{eqnarray}
Note that the kinetic term is still diagonal, i.e. there are no cross terms involving $\mathbf{p}_i\mathbf{p}_j$ with $i\neq j$.
Next, we add and subtract the term 
\begin{eqnarray} 
\Ha_\pm = \sum_{i=1}^{N-1}\frac{G m'_i M_{i} }{|\mathbf{r}'_i|}. \label{eq:hpm}
\end{eqnarray}
After grouping terms in the Hamiltonian, we arrive at
\begin{eqnarray}
\mathcal{H} &=& 
\underbrace{\frac{{\mathbf{p}'_0}^2}{2m'_0}\vphantom{\sum_{i=1}^{N-1} \frac{{\mathbf{p}'_i}^2}{2m'_i} }}_{\mathcal{H}_{0}} 
+\underbrace{\sum_{i=1}^{N-1} \frac{{\mathbf{p}'_i}^2}{2m'_i} -\sum_{i=1}^{N-1}\frac{G m'_i M_{i} }{|\mathbf{r}'_i|}}_{\mathcal{H}_{\rm Kepler}} \nonumber \\
&&  + \underbrace{\sum_{i=1}^{N-1}\frac{G m'_i M_{i} }{|\mathbf{r}'_i|}- \sum_{i=0}^{N-1} \sum_{j=i+1}^{N-1} \frac{Gm_im_j}{|\mathbf{r}_i-\mathbf{r}_j|}}_{\mathcal{H}_{\rm Interaction}}.\label{eq:hamsplit}
\end{eqnarray}
The first term, $\mathcal{H}_0$, simply describes the motion of the centre-of-mass $\mathbf{r}'_0$ along a straight line.
For that reason this term is often ignored.
However, we keep it which will allow us to integrate particles without any restriction to a particular frame of reference. 

The terms $\mathcal{H}_{\rm Kepler}$ can be split up further into a sum of 
\begin{eqnarray}
\left(\mathcal{H}_{\rm Kepler}\right)_i &=& \frac{{\mathbf{p}'_i}^2}{2m'_i} - \frac{G m'_i M_{i} }{|\mathbf{r}'_i|}. \label{eq:HKepler}
\end{eqnarray}
Each of the Hamiltonians $\left(\mathcal{H}_{\rm Kepler}\right)_i$ describes the Keplerian motion of the $i$-th particle with mass $m_i$ around the centre-of-mass of all interior particles with total mass $M_{i-1}$.

After some more algebra, the interaction term can be simplified and split into two parts, one of which can be easily computed in Jacobi coordinates and the other in Cartesian coordinates of the inertial frame
\begin{eqnarray}
\Ha_{\rm Interaction} = 
\sum_{i=2}^{N-1} \frac{ G{m'}_i M_i }{|\mathbf{r}'_i|} 
 - \sum_{i=0}^{N-1} \sum_{\substack{j=i+1\\j\neq 1}}^{N-1} \frac{Gm_im_j}{|\mathbf{r}_i-\mathbf{r}_j|}.
\end{eqnarray}

One important point to note is that our choice of $\Ha_\pm$ is slightly different from that used by \cite{solarsystemdynamics} and \cite{WisdomHolman1991}.
These authors use 
\begin{eqnarray}
\left(\Ha_\pm\right)_{\rm WH1991} &=& \sum_{i=1}^{N-1}\frac{G m'_i \, \mathcal{M}_i }{|\mathbf{r}'_i|}.
\end{eqnarray}
where $\mathcal{M}_i = m_0 \frac{M_i}{M_{i-1}}$.
Their choice leads to the usual disturbing function in perturbation theory.
We conducted various tests but found no significant difference between these mass choices. 
We therefore chose our prescription, Eq.\:\ref{eq:hpm}, which has a simpler physical interpretation: the mass entering Kepler's third law is simply the interior mass.

%%%%%%%%%%%%%%%%%%%%%%%%%%%%%%%%%%%%%%%%%%%%%%%%%
\subsection{Wisdom-Holman Mapping}
Our goal is to find a solution to the equations of motion for particles governed by the Hamiltonian in Eq.~\ref{eq:H}.
No analytic solution exists to the full Hamiltonian and we thus need to find an approximate solution. 
There are many different ways to do that.
Here, we describe the idea of constructing a symplectic integrator by means of splitting the Hamiltonian into smaller parts, each of which can be easily integrated.

The introduction of Jacobi coordinates led us to the Hamiltonian splitting described in Sect.~\ref{sec:splitting}. 
Analytic solutions can be found for the evolution of the system under each of the individual Hamiltonians $\Ha_0$ and $\Ha_{\rm Interaction}$.
The solution to $\Ha_0$ simply corresponds to motion along a straight line.
The solution to $\Ha_{\rm Interaction}$ is a kick step where the velocities change due the inter-particle accelerations but the position remain constant.
The solution to $\Ha_{\rm Kepler}$ is a set of two-body Kepler orbits, which can also be easily solved with an iterative algorithm. 
We discuss the details related to the Kepler problem in Sect.~\ref{sec:kepler}.

Now that we have broken down the full Hamiltonian into individual Hamiltonians, to all of which we know the solution (or can easily calculate it), we can construct a symplectic integrator for the total Hamiltonian using an operator split method \citep[e.g.][]{SahaTremaine1992}.
Let us describe the evolution of particles under a Hamiltonian $\Ha$ for a time $\mathit{dt}$ using the operator notation $\hat{\Ha}(\mathit{dt})$.
The notation $\hat{\Ha}_2(\mathit{dt})\circ\hat{\Ha}_1(\mathit{dt})$ means applying operator $\hat{\Ha}_1$ first, then applying operator $\hat{\Ha}_2$.
It is easy to see that many of the operators commute with each other, i.e.
\begin{eqnarray}
\left[\hat \Ha_0, \hat \Ha_{\rm Kepler} \right] &=& 0 \\
\left[\hat \Ha_0, \hat \Ha_{\rm Interaction} \right] &=& 0\\
\left[\left(\hat \Ha_{\rm Kepler}\right)_i , \left(\hat \Ha_{\rm Kepler}\right)_j \right] &=& 0\quad \forall i,j,
\end{eqnarray}
where $[\hat{\Ha}_1,\hat{\Ha}_2] = \hat{\Ha}_1\circ \hat{\Ha}_2- \hat{\Ha}_2\circ \hat{\Ha}_1$.
This leads to the following Drift-Kick-Drift (DKD) operator splitting scheme, which we refer to as the Wisdom-Holman map:
\begin{enumerate}[labelwidth=1.5cm,labelindent=10pt,leftmargin=1.2cm]
\item [(Drift)] Evolve the system under $\hat \Ha_{\rm Kepler}(\mathit{dt}/2) \circ \hat\Ha_0(\mathit{dt}/2)$. 
\item [(Kick)] Evolve the system under $\hat \Ha_{\rm Interaction}(\mathit{dt})$. 
\item [(Drift)] Evolve the system under $\hat \Ha_{\rm Kepler}(\mathit{dt}/2) \circ \hat\Ha_0(\mathit{dt}/2)$. 
\end{enumerate}
The ordering of $\hat \Ha_{\rm Kepler}$ and $\Ha_0$ in the first and last step doesn't matter as they commute.
The first and last steps can be combined if the system is evolved for multiple timesteps.

Note that the evolution of $\Ha_{\rm Kepler}$ and $\Ha_0$ is most easily accomplished in Jacobi coordinates.
The interaction Hamiltonian $\Ha_{\rm Interaction}$, however, contains terms that depend on both the Cartesian and Jacobi coordinates. 
The simplest way to calculate these terms is to convert to Cartesian coordinates, evaluate the $\mathbf{r}_i-\mathbf{r}_j$ term, convert the accelerations back to Jacobi accelerations, and calculate the remaining terms.

%%%%%%%%%%%%%%%%%%%%%%%%%%%%%%%%%%%%%%%%%%%%%%%%%%%%%%%%%%%%
\subsection{Symplectic Correctors}\label{sec:correctors}
The operator splitting method used in the symplectic integrator discussed above effectively adds high frequency terms to the Hamiltonian.
An argument often used in favour of symplectic integrators is that, although these high-frequency terms alter the Hamiltonian, they do not change the long term evolution as they average out. 
However, they do lead to relatively large short term oscillations, for example in the energy error.

The idea of a symplectic corrector, first used by \cite{TittemoreWisdom1989} and fully developed by \cite{Wisdom1996}, is to remove some of these high frequency terms using perturbation theory. 
The basic procedure is as follows.
Before the start of an integration, we convert from real coordinates to so-called mapping coordinates.
Then we perform the integration using our standard symplectic map.
After the simulation has finished (or whenever we need an output) we convert back from mapping to real coordinates.
The symplectic corrector operator that we use is a combination of several $\hat \Ha_{\rm Interaction}(\mathit{dt})$ and $\hat \Ha_{\rm Kepler}(\mathit{dt}) \circ \hat\Ha_0(\mathit{dt})$ operators applied for different (positive and negative) intervals $\mathit{dt}$.
If $\epsilon$ is the order of the perturbations, i.e. the mass ratio and therefore the relative magnitude of $\hat \Ha_{\rm Interaction}$ compared to $\hat \Ha_{\rm Kepler}$, then one can show that the use of symplectic correctors can lead to a scheme of order $O(\epsilon \mathit{dt}^{K} ) + O(\epsilon^2 \mathit{dt}^2)$ where $K$ is the order of the symplectic corrector \citep{MikkolaPalmer2000}.
A second order Wisdom-Holman map without symplectic correctors has an energy error of order $O(\epsilon^2 \mathit{dt}^2)$.
Because this coordinate transformation for the symplectic corrector is only performed for outputs and at the beginning and end of the simulation, its effect on the speed of the algorithm is negligible for sparse output.

A full derivation of the symplectic correctors would go beyond the scope of this paper and we refer the reader to \cite{Wisdom1996} and \cite{MikkolaPalmer2000}.
The corrector coefficients are listed in a compact form in \cite{Wisdom2006}.

We implement a third, fifth, seventh and eleventh order symplectic corrector for \whf.
Whether the high order-symplectic correctors provide any improvement over the low-order ones depends on the mass ratios in the system.
For Jupiter-mass planets, a symplectic corrector of fifth order is no less accurate than a higher order one.
If in doubt, there is no harm done in using a higher-order corrector as the speed implications are minimal.
Thus, we implement the eleventh-order symplectic corrector by default.

%%%%%%%%%%%%%%%%%%%%%%%%%%%%%%%%%%%%%%%%%%%%%%%%%%%%%%%%%%%%

\subsection{Chaos Indicators}
A powerful tool for studying the long term evolution of Hamiltonian systems is the Lyapunov characteristic number (LCN). 
The inverse of the LCN is the Lyapunov timescale and gives an estimate of how fast two nearby particle trajectories diverge.
If the system is chaotic, the divergence is exponential in time and the Lyapunov timescale is finite.
Thus, measuring the LCN gives us an estimate of whether the system is chaotic and, if so, on what timescale.

A more recent approach with similar informative value is the Mean Exponential Growth factor of Nearby Orbits, or MEGNO for short \citep{Cincotta2003}. 
The MEGNO, $Y(t)$, is a scalar function of time, and provides a clear picture of resonant structures and of the locations of stable and unstable periodic orbits.

There are two ways to calculate the LCN or the MEGNO. 
Conceptually the simplest is to integrate an additional shadow particle for each body in the simulation, i.e. a particle with slightly perturbed initial conditions.
One can then directly measure the divergence of each particle's path from its shadow.
The second approach is to consider each body's six-dimensional displacement vector $\pmb{\delta_i}$ from its shadow (in both position and velocity) as a dynamical variable.
One can then obtain differential equations for each $\pmb{\delta_i}$ vector by applying a variational principle to the trajectories of the original bodies.
We choose to follow the latter approach, as it is both faster and numerically more robust \citep{Tancredi2001}.
In this scheme, one can imagine shadow particles with phase-space coordinates $\pmb \xi^s_i = \pmb\xi_i + \pmb\delta_i$, where $\pmb\xi_i=(\mathbf{r_i}, \mathbf{v_i})$ is the phase-space coordinate of the $i$-th original particle. 
Initially, we set each component of $\pmb \delta_i$ to a small value.

We follow the work of \cite{MikkolaInnanen1999} who describe how to efficiently couple the variational equations to the original equations of motion.
This allows us to construct a symplectic integrator for the variational equations (a symplectic tangent map). 
An important advantage of this method is that we only solve Kepler's equation once for each particle/shadow-particle pair (one of the most time-consuming steps in a Wisdom-Holman integrator for small particle numbers).

The MEGNO is then straightforwardly computed from the variations as \citep{Cincotta2003}
\begin{eqnarray}
Y(t) = \frac{2}{t} \int_0^t  t' \frac{\sum_{i=0}^{N-1} \dot{\pmb{\delta}_i}(t') \cdot \pmb{\delta}_i(t')}{\sum_{i=0}^{N-1} \pmb{\delta}_i^2(t')} dt'.
\end{eqnarray}
If $Y(t)\rightarrow \infty$, then the system is chaotic.
For quasi-periodic orbits, the MEGNO converges to a finite value, $Y(t)\rightarrow 2$  \citep[e.g.][]{Hinse2010}.

One can obtain the Lyapunov characteristic number (LCN), the inverse of the Lyapunov timescale, from the time evolution of the MEGNO via a linear least square fit to~$Y(t)$. 

%%%%%%%%%%%%%%%%%%%%%%%%%%%%%%%%%%%%%%%%%%%%%%%%%%%%%%%%%%%%

\subsection{Kepler Problem with Variations}\label{sec:kepler}
In this section, we summarize how to solve the two-body Kepler problem numerically, including the variational equations.
Although the solution has been known since the days of Newton, the transcendental nature of Kepler's equation does not admit a closed-form mathematical expression.

We closely follow the work of \citep{MikkolaInnanen1999} where the reader can find additional information that we have left out.
The equivalent one-body Hamiltonian for the Kepler problem is
\begin{eqnarray}
H_{\rm Kepler} &=& \frac12 \mathbf{v}^2 - \frac{M}{|\mathbf{r}|},
\end{eqnarray}
where $M$ is the total mass of the two bodies.  
For consistency with \cite{MikkolaInnanen1999}, we have dropped the primes, have scaled out $m'_i$ from $p'_i$, and rewritten Eq.~\ref{eq:HKepler} in non-dimensional form, i.e. the gravitational constant $G=1$ for the remainder of this paper.
However, we have taken care to remove any dependence on the choice of units from our implementation, so $G$ can be freely set by the user in our implementation of the algorithm.

Our task is to find the final positions and velocities $\mathbf{r}$ and $\mathbf{v}$ of a particle evolving under this Hamiltonian for some time $\mathit{dt}$, given the initial conditions $\mathbf{r}_0$ and $\mathbf{v}_0$.
Thus, we seek the effect of the operator $\hat H_{\rm Kepler}(\mathit{dt})$.

It is advantageous to solve the Kepler problem numerically using the Gauss f and g functions, which express the relevant quantities in terms of $\mathbf{r}_0$ and $\mathbf{v}_0$ \citep{WisdomHolman1991}. This avoids the computationally expensive conversion between Cartesian and classical orbital elements, and avoids coordinate singularities associated with circular orbits. We find that it is advantageous to use universal variables in this solution \citep{Stumpff1962}. This approach provides greater speed and numerical stability compared to a solution using elliptic elements. It also avoids the singularity associated with the transition from elliptic to hyperbolic motion.

To solve the analogue of Kepler's equation for the particle's position in time, we make use of several special functions.
Let us begin by defining the $c$-functions \citep{Stumpff1962} as a series expansion:
\begin{eqnarray}
c_n(z) \equiv \sum_{j=0}^\infty \frac{(-z)^j}{(n+2j)!} \label{eq:c},
\end{eqnarray}
which satisfy the recursion relation
\begin{eqnarray}
c_n(z) &=& \frac{1}{n!} - z \,c_{n+2}. \label{eq:recur}
\end{eqnarray}
The $c$-functions are related to trigonometric functions, for example
\begin{eqnarray}
c_0(z) = \cos \sqrt{z} \quad\quad \text{and}\quad\quad c_1(z) = \frac{\sin\sqrt{z}}{\sqrt{z}},
\end{eqnarray}
and thus satisfy the following relationships \citep{Mikkola1997}, which are related to the half-angle formula for trigonometric functions:
\begin{eqnarray}
c_5(z) &=& \frac1{16} \left[ c_5(z/4) + c_4(z/4) + c_3(z/4)c_2(z/4)\label{eq:crel5}\right]\\
c_4(z) &=& \frac1{8}  c_3(z/4) \left[1 + c_1(z/4).\label{eq:crel4}\right]
\end{eqnarray}
Values for $c_0$ through $c_3$ are then readily computed from Eq.~\ref{eq:recur}.
Next, we introduce the so called $G$-functions \citep{StiefelScheifele1971} which in turn depend on the $c$-functions:
\begin{eqnarray}
G_n(\beta,X) \equiv X^n c_n(\beta X^2). \label{eq:G}
\end{eqnarray}
The $G$-functions also satisfy recursion relationships similar to those mentioned above for the $c$-functions.
We can easily calculate derivatives of $G_n$ by looking at the series expansion of $c_n$ \citep[see][for details]{MikkolaInnanen1999}.
With this framework, we can now write down the steps needed to find the solution to the Kepler Hamiltonian in compact form.

First, we need to calculate the following three quantities from the initial conditions $\mathbf{r}_0, \mathbf{v}_0$:
\begin{eqnarray}
\beta &=& \frac{2M}{r_0}-v_0^2\\
\eta_0 &=& \mathbf{r}_0 \cdot \mathbf{v}_0 \\
\zeta_0 &=& M-\beta r_0
\end{eqnarray}
where $r_0 = |\mathbf{r}_0|$ and $v_0 = |\mathbf{v}_0|$. Note that the semi-major axis $a$ can be written as $a=M/\beta$. 

Second, we need to solve Kepler's equation which, using the above notation, takes the form
\begin{eqnarray}
r_0 X + \eta_0 G_2(\beta,X) + \zeta_0 G_3(\beta,X) - \mathit{dt} = 0. \label{eq:kepler} 
\end{eqnarray}
We solve this equation for $X$.
This is a non-algebraic (i.e. transcendental) equation that we need to solve iteratively, for example using Newton's method.
In Sect.~\ref{sec:newton}, we describe our algorithm in detail.

Third, having solved Kepler's Equation, we can calculate the so called Gau{\ss} $f$ and $g$-functions as well as their time derivatives via
\begin{align}
f &= 1 - M \frac{G_2}{r_0}\quad & \dot f &= -\frac{M\,G_1}{r_0r} \label{eq:fg}&\\
g &= dt - M G_3            & \dot g &= 1-\frac{M\,G_2}{r},\label{eq:fg2}
\end{align}
where $r = r_0 + \eta_0 G_1 + \zeta_0 G_2$.
Note that all the $G$-functions depend on $\beta$ and the $X$ value found in the second step.

Fourth, we write the final positions and velocities as a linear transformation of the initial conditions using the Gau{\ss} $f$ and $g$-functions:
\begin{align}
\mathbf{r} &= f \mathbf{r}_0 + g \mathbf{v}_0
&\mathbf{v} &= \dot f \mathbf{r}_0 + \dot g \mathbf{v}_0.\label{eq:fgupdate}&&
\end{align}
This completes the solution of the Kepler problem.

To solve for the variational equations, we also make use of the $G$-functions.
Fortunately, we only need to solve Kepler's equation once (to solve for $X$). 
We then get the solution for the variational equations without solving another transcendental equation and thus have only one iteration loop per timestep for both the particle and its variational counterpart.
The position and velocity components of $\pmb \delta$ at the end of the timestep,~$\delta \mathbf{r}$ and $\delta \mathbf{v}$, can be written as
\begin{eqnarray}
\delta \mathbf{r} &=& f \;\delta \mathbf{r}_0 + g \;\delta\mathbf{v}_0 + \mathbf{r}_0 \;\delta f+ \mathbf{v}_0 \;\delta g\label{eq:var1}\\
\delta \mathbf{v} &=& \dot f \;\delta \mathbf{r}_0 + \dot g \;\delta\mathbf{v}_0 + \mathbf{r}_0 \;\delta \dot f+ \mathbf{v}_0 \;\delta \dot g,\label{eq:var2}
\end{eqnarray}
where the variations $\delta f$, $\delta g$, $\delta \dot f$ and $\delta \dot g$ can be derived from Eqs.~\ref{eq:fg}-\ref{eq:fg2} (see \citealt{MikkolaInnanen1999} for the explicit expressions).

%%%%%%%%%%%%%%%%%%%%%%%%%%%%%%%%%%%%%%%%%%%%%%%%%
\subsection{Types of Numerical Errors}\label{sec:errors}
There are three distinct effects contributing to the energy error of a symplectic integrator \citep[see e.g.,][]{QuinnTremaine1990}.
See also \cite{ReinSpiegel2015} for a similar discussion for non-symplectic integrators.

First, there is an error term associated with the integrator itself because we are not solving the equations of motion for the Hamiltonian $\mathcal{H}$ exactly.
For symplectic integrators such as those discussed here, this error term is bound and we call it $E_{\rm bound}$.
If the mass ratio of the planets to the star is $\epsilon$, then the order of this error term is roughly $O(\epsilon\, \mathit{dt}^{2})$ for integrators without symplectic corectors and  $O(\epsilon \,\mathit{dt}^{K} ) + O(\epsilon^2 \,\mathit{dt}^2)$ for those with symplectic correctors (see Sect.~\ref{sec:correctors}).
Note that $E_{\rm bound}$ is independent of time $t$.

Second, there is an error term associated with the finite precision of numbers represented on a computer. 
We can only represent a small subset of all real numbers exactly in floating-point precision.
Thus after every operation such as an addition or multiplication, the computer rounds to a nearby floating-point number.
For CPUs and compilers that follow the IEEE754 standard \citep{IEEE754}, we are guaranteed to round to the nearest floating-point number.
Thus, if all operations follow the IEEE754 standard, then as long as the algorithm itself is unbiased, we expect the error to grow as the square root of the number of operations, i.e.  $E_{\rm rand} \sim \sqrt{N}\sim \sqrt{t}$, where $N$ is the number of timesteps.
This is the best behaviour achievable; to do better we would have to move to extended precision or use fewer operations.
This fundamental limit is known as Brouwer's law~\citep{Newcomb1899,Brouwer1937}.

Third, if any parts of the integration algorithm are biased, the errors will be correlated.
This leads to a faster long-term energy-error growth than if errors are uncorrelated; it grows linearly with time, i.e. $E_{\rm bias}\sim N \sim t$.

For a given integrator, which of these three error terms dominates depends on the nature of the simulation, the timestep, and the total integration time (number of timesteps).

%%%%%%%%%%%%%%%%%%%%%%%%%%%%%%%%%%%%%%%%%%%%%%%%%
%%%%%%%%%%%%%%%%%%%%%%%%%%%%%%%%%%%%%%%%%%%%%%%%%
%%%%%%%%%%%%%%%%%%%%%%%%%%%%%%%%%%%%%%%%%%%%%%%%%
\section{Improvements}\label{sec:improvements}
The algorithms we describe in Sect.~\ref{sec:background} have been used successfully for many years.
In the following, we show how to significantly improve the speed and accuracy of the algorithms by taking special care in the implementation of several details, many of which are related to finite floating-point precision on modern computers.

For the remainder of this paper, we will assume that we work with a CPU that follows the IEEE~754 standard for floating-point arithmetic.
Most importantly, we assume that all floating-point operations follow the \textit{rounding to nearest, ties to even} rule \citep{IEEE754}.
What follows is in principle applicable to any precision.
However, we work exclusively in double floating-point precision (64 bit) which is used on almost all modern CPUs.

%%%%%%%%%%%%%%%%%%%%%%%%%%%%%%%%%%%%%%%%%%%%%%%%%%%%%%%%%%%%
\subsection{Jacobi Coordinate Transformations}\label{sec:jacobi}
The evolution under the effect of the interaction Hamiltonian is most efficiently done in Cartesian coordinates. 
On the other hand, the evolution of the Kepler Hamiltonian is easier in Jacobi coordinates. 
We thus need an efficient way to convert to and from Jacobi coordinates.

Luckily, the conversion from Cartesian to Jacobi coordinates and back can be done efficiently in $\mathcal{O}(N)$.
We construct the algorithms from the definitions above and list them here in pseudo code.
As before, primes denote Jacobi coordinates,
Note that these algorithms work even if some of the bodies are test particles with $m_i=0$ (for $i\neq 0$).
To convert from Cartesian to Jacobi coordinates:\vspace{0.2cm}
\begin{algorithmic}
	\State $\mathbf{R} \gets m_0 \cdot \mathbf{r}_0$
	\For{$i \gets 1,N-1$}
		\State $\mathbf{r}'_i \gets \mathbf{r}_i - \mathbf{R}/M_{i-1}$
		\State $\mathbf{R}  \gets \mathbf{R}  \cdot (1 +m_i/M_{i-1}) + m_i\cdot\mathbf{r}'_i$ 
	\EndFor
	\State $\mathbf{r}'_0 \gets \mathbf{R} / M_{N-1}$ \Comment{This is the centre-of-mass.}
\end{algorithmic}\vspace{0.2cm}
Similarly, we construct the algorithm to convert back from Jacobi to Cartesian coordinates as follows:\vspace{0.2cm}
\begin{algorithmic}
	\State $\mathbf{R} \gets \mathbf{r}'_0\cdot M_{N-1}$ \Comment{Centre of mass.}
	\For{$i \gets N-1,1$} \Comment{Loop is in reverse order.}
		\State $\mathbf{R} \gets (\mathbf{R}- m_i \cdot \mathbf{r}'_i) / M_i$
		\State $\mathbf{r}_i \gets \mathbf{r}'_i + \mathbf{R}$ 
		\State $\mathbf{R}  \gets \mathbf{R} \cdot M_{i-1}$ 
	\EndFor
	\State $\mathbf{r}_0 \gets \mathbf{R}/m_0$\Comment{Setting the coordinate of the 0-th particle.}
\end{algorithmic}\vspace{0.2cm}
We thoroughly tested the conversions to and from Jacobi coordinates to ensure they are unbiased. 
This task turns out to be much harder than we na{\"i}vely expected.
As an example, consider the following algorithm which is formally equivalent to the above but numerically much less stable.\vspace{0.2cm}
\begin{algorithmic}
	\State $\mathbf{R} \gets 0$
	\For{$i \gets N-1,1$} \Comment{Loop is in reverse order.}
		\State $\mathbf{r}_i \gets \mathbf{r}'_0 + M_{i-1}/M_i \cdot \mathbf{r}'_i -\mathbf{R}$ \Comment{$\mathbf{r}'_0$ is the centre-of-mass.}
		\State $\mathbf{R}  \gets \mathbf{R}  + m_i/M_i \cdot \mathbf{r}'_i$ 
	\EndFor
	\State $\mathbf{r}_0 \gets \mathbf{r}'_0 - \mathbf{R}$\Comment{Setting the coordinate of the 0-th particle.}
\end{algorithmic}\vspace{0.2cm}
In the above algorithm, we access $\mathbf{r}_0'$ multiple times and have to do a subtraction in the last step. 
This significantly promotes error propagation and leads to floating-point errors that can be orders of magnitudes higher than in the other implementation.
After many timesteps, this leads to a linear secular growth in the energy error.

%%%%%%%%%%%%%%%%%%%%%%%%%%%%%%%%%%%%%%%%%%%%%%%%%%%%%%%%%%%%
\subsection{Implementation of Newton's Method}\label{sec:newton}
To solve Kepler's equation (Eq.~\ref{eq:kepler}) for $X$, we need to use an iterative scheme.
We now describe our implementation of Newton's method in floating-point arithmetic.
The straightforward implementation is an iteration loop that terminates when the change to $X$ is small, e.g., \vspace{0.2cm}
\begin{algorithmic}
	\State $X \gets initial\; guess$
	\Repeat
		\State $dX \gets -f(X)/f'(X)$
		\State $X \gets X + dX$
	\Until{ $|dX/X| < \eps$}.
\end{algorithmic} \vspace{0.2cm}
Here, $\eps$ is a small number just above machine precision, typically~$\eps \sim10^{-15}$.
We use a different implementation of Newton's method that is both faster and more accurate, despite the fact that it is algebraically equivalent to the above implementation. \vspace{0.2cm}
\begin{algorithmic}
	\State $X \gets initial\; guess$
	\State $X_{\rm prev 1} \gets \mathit{NaN}$\Comment{Any number different from $X$ works.}
	\Repeat
		\State $X_{\rm prev 2} \gets X_{\rm prev 1}$
		\State $X_{\rm prev 1} \gets X$
		\State $X \gets (X\cdot f'(X)-f(X))/f'(X)$
	\Until{$X = X_{\rm{ prev 1}}$ or $X = X_{\rm{ prev 2}}$ } %\Comment{evaluated in FP precision}
	%\Until{$X = X_{\rm old}$}
\end{algorithmic} \vspace{0.2cm}
Note that the equal sign in the above breakout condition is evaluated in floating-point precision.
In comparison to the first algorithm, at each iteration step we test whether the iteration has converged by a simple comparison rather than by a slow division and absolute-value operation.

We keep track of two previous values instead of just one because for certain initial conditions, the iteration can cycle indefinitely between two nearby floating-point numbers and not converge to a single floating-point number.

Our implementation thus ensures that the value of $X$ is more accurately calculated than in the straightforward implementation using a heuristic value of $\eps$.
A further advantage of rewriting Newton's method in the above form is that the term on the right-hand-side of the last line can be simplified significantly for the Kepler problem, giving:
\begin{eqnarray}
X \gets  \frac{X (\eta_0 G_1 + \zeta_0 G_2) -\eta_0 G_2 -\zeta_0 G_3+dt }{r_0+\eta_0 G_1 +\zeta_0 G_2} \label{eq:kepeq}
\end{eqnarray} 
where the $G$'s on the right-hand-side all depend on $X$ and $\beta$ (see Eq.~\ref{eq:G}).

We also experimented with higher-order generalizations of Newton's method (Householder's methods).  For typical cases where the orbits are not extremely elliptical ($e\lesssim 0.99$) and the timestep is much smaller than the shortest orbital period, we found Newton's method to always be fastest.  
This is because when the value and derivatives of the function are easily evaluated, the precision gain from these higher-order methods does not compensate for the increased computation cost of each iteration.
In other words, while higher-order methods will converge in fewer iterations than Newton's method, the overall computation time is longer.  
At large eccentricities and long timesteps, the $G$-function evaluations become expensive (one must recursively apply the quarter-angle formulas described in Sect.~\ref{sec:cs}), and higher-order methods are helpful.  
For large eccentricities we use a higher order method described in detail in Sect.~\ref{sec:largeE}.  
To safeguard against rare cases where Newton's method might fail, we also implemented a failsafe bisection method.  
We find that the bisection method is only triggered when the timestep is comparable to the orbital period.

%%%%%%%%%%%%%%%%%%%%%%%%%%%%%%%%%%%%%%%%%%%%%%%%%%%%%%%%%%%%
\subsection{The Initial Guess for Kepler's Equation:  Short Timesteps}
The quantity X in Eq.~\ref{eq:kepeq} can also be expressed as
\begin{eqnarray}
X = \int_{t_0}^{t_0 + \mathit{dt}} \frac{\mathit{dt}'}{r} = \mathit{dt} \cdot{\langle r^{-1}\rangle} \label{eq:xint}
\end{eqnarray}
where $t_0$ is the time at the beginning of the timestep, and $\langle r^{-1}\rangle$ is the time-averaged value of $r^{-1}$ over the interval $[t_0, t_0+\mathit{dt}]$.  
Thus, if the orbit's eccentricity $e$ is low, or more generally if the timestep is short enough that the orbital radius does not vary much, then $X \approx \mathit{dt}/r_0$.  
The troublesome cases are highly eccentric orbits near pericentre where the radius changes rapidly.  
For such cases, the radius varies by a factor of $1+e \approx 2$ from pericentre to a true anomaly of $90^\circ$.  
We can therefore estimate the timescale over which the orbital radius varies near pericentre as
\begin{eqnarray}
T_{char} = \frac{q}{v_q} = \frac{a(1-e)}{na}\Bigg(\frac{1-e}{1+e}\Bigg)^{1/2} \sim \frac{(1-e)^{3/2}}{n}, \label{eq:tchar}
\end{eqnarray}
where $q$ is the pericentre distance, $v_q$ is the speed at pericentre and $n$ is the mean motion.  
Thus, if one does not resolve pericentre passages (i.e., $n\,\mathit{dt} = \Delta M \gtrsim (1-e)^{3/2}$), $X$ will differ from $\mathit{dt}/r_0$ near pericentre (but may nevertheless conform to the simple approximation at apocentre where the body moves slowly).  

More quantitatively, one can non-dimensionalize Eq.~\ref{eq:kepeq}, setting $\tilde{X} = r_0 X / \mathit{dt}$.  
One can then solve the equation perturbatively, assuming the deviations from $\tilde{X} = 1$ are small.  
This procedure requires that the following three non-dimensional parameters in the equation also be much smaller than unity, 
\begin{align}
\chi \equiv \frac{\beta \mathit{dt}^2}{r_0^2}\quad \eta \equiv \frac{\eta_0\mathit{dt}}{r_0^2} \quad \zeta \equiv \frac{\zeta_0\mathit{dt}^2}{r_0^3}.
\end{align}
One can show that when our heuristic estimate $\Delta M \ll (1-e)^{3/2}$ is satisfied, $\chi$, $\eta$, $\zeta \ll 1$.  
In this case, one can extend the solution of Eq.~\ref{eq:kepeq} to higher order.  
For the initial guess in our algorithm, we go up to second order
\begin{eqnarray}
X = \frac{\mathit{dt}}{r_0} \cdot \left(1 - \frac12 \eta \right). \label{eq:xinit}
\end{eqnarray}
We experimented with higher-order initial guesses (see \citealt{Danby1987} for explicit expressions), but found these to be slower, even for small eccentricities and timesteps.  
This can again be attributed to the computational efficiency of each iteration of Newton's method.

%%%%%%%%%%%%%%%%%%%%%%%%%%%%%%%%%%%%%%%%%%%%%%%%%%%%%%%%%%%%
\subsection{Large Eccentricities and Timesteps} \label{sec:largeE}
The previous two sections describe an optimized algorithm for solving Kepler's equation when the timestep and eccentricities are low.  
We have also developed an improved handling of high-eccentricity/long-timestep cases.  
In this regime, both the solver and initial guess should be modified.

Like previous authors \citep{Conway1986, Danby1987}, we found the root-finding method of Laguerre-Conway to be most stable.  
However, unlike \cite{Danby1987}, who finds the method to always converge (presumably using comparatively small timesteps), we often have to resort to bisection when the timestep is comparable to the orbital period.  
Of course, such long timesteps should not be chosen anyway, since they poorly sample inter-planet interactions, and are more susceptible to timestep resonances \citep{WisdomHolman1992,ToumaWisdom1993,Rauch1999}.

We also had to modify the breakout condition used for Newton's method.  
While the Laguerre-Conway algorithm sometimes also bounces between two floating-point values once it has converged, in this regime the method often executes larger-period cycles (e.g., it will periodically repeat the last eight floating-point numbers).  
We therefore chose to store the values from each iteration and exit the loop whenever a result was repeated.

One way to determine which solver should be used is to check whether $\mathit{dt}$ is smaller than $T_{char}$ (Eq.~\ref{eq:tchar}).  
However, because $T_{char}$ is expensive to compute from $\mathbf{r}_0$ and $\mathbf{v}_0$, we instead check how much the first iteration of Newton's method deviates from the initial guess, as a fraction of $2\pi \beta^{-1/2}$.  
The latter is a natural quantity to compare against since it is the value of $X$ when the timestep is equal to the orbital period.  
We found a threshold of $\sim 1\%$ to strike a good balance over a wide parameter range in timestep/eccentricity space, though the algorithm's speed is not particularly sensitive to the exact value adopted.

Finally, the method can be sped up in this regime with an improved initial guess for $X$, since $\mathit{dt}/r_0$ in Eq.~\ref{eq:xinit} blows up near pericentre as the eccentricity gets large.  
\cite{Danby1983} provide a widely used initial guess using classical orbital elements but, to our knowledge, no comparably simple initial guess has been found for universal variables.  

In this high-eccentricity / long timestep regime, most existing methods using universal variables choose to make the expensive conversion to orbital elements and use Danby's guess.  
We instead observed that because $\langle r^{-1} \rangle = a^{-1}$ over one orbital period, $X = \mathit{dt}/a$ for a timestep of one orbit.  
We find that over a relevant parameter range with timesteps logarithmically spaced between 0.03 and 1 orbital periods, and eccentricities between 0.999 and 0.9999, our improved guess is faster than converting to orbital elements and using Danby's by $\approx 30\%$.
In a manner analogous to that described in the previous section, we also solved Eq.~\ref{eq:kepeq} perturbatively around $X = \mathit{dt}/a = \beta\; \mathit{dt} / M$ in the regime $\chi \gg 1$, but we found the second-order solution to be a slower initial guess than the simple $X = \beta \; \mathit{dt} / M$.  

%%%%%%%%%%%%%%%%%%%%%%%%%%%%%%%%%%%%%%%%%%%%%%%%%
\subsection{Implementation of $c$-functions} \label{sec:cs}
Finding a solution to Kepler's equation is done iteratively and is thus the most expensive step in solving the Kepler problem.
The iteration itself involves the calculation of multiple $G$-functions, which in turn require the calculation of $c$-functions.
Thus, it is particularly important to optimize these functions for both speed and accuracy. 
When calculating chaos indicators, we need~$c_0$, $c_1$, $c_2$, $c_3$, $c_4$ and $c_5$. 
If we are not integrating the variational equations, we only need $c_0$, $c_1$, $c_2$ and $c_3$.

We first ensure that $z$ is smaller than~$0.1$ to guarantee that the series expansion of $c$  in Eq.~\ref{eq:c} converges.
We do this by dividing $z$ repeatedly by 4. 
Note that divisions by powers of 2 are fast and exact in floating-point arithmetic.
To calculate the series expansion, we need an inverse factorial for every term. 
Calculating this inverse factorial by multiplying floating-point numbers and then implementing a floating-point division would be very slow.
We found that the fastest way to calculate the inverse factorial is to use a simple lookup table. 
We checked that the series expansions of the $c$-functions converge very quickly for small $z$ and thus we only store inverse factorials up to $1/34!$ in the lookup table.
Any larger factorial would contribute less than one part in $10^{16}$ to the sum and can thus be neglected (as we work in double floating-point precision).

We always calculate the first two terms in the series expansion. 
We then enter a loop and add more terms until the result no longer changes.
Because $z$ is small and the inverse factorials decrease quickly, we are assured that the series will converge to a single floating-point number.
This allows us to simply check whether the value changes from one iteration to the next, which is much faster than evaluating relative changes (cf. Sect.~\ref{sec:newton}). 
 
Once the $c$-functions are calculated for the small $z$ value, we use the relations in Eqs.~\ref{eq:crel5}-\ref{eq:crel4} with Eq.~\ref{eq:recur} to calculate the $c$-functions for the original $z$ value. 

Because this algorithm is an integral part of the integrator, we list the function to calculate $c(z)$ in pseudo code: \vspace{0.2cm}
\begin{algorithmic}
	\State $n \gets 0$		\Comment{Counter for quarter-angle formula.}
	\While{$z>0.1$}			\Comment{Ensure that $z$ is small.}
		\State $z \gets z/4$ 	
		\State $n \gets n+1$
	\EndWhile
	\State $c_4 \gets \frac{1}{4!} - z \cdot \frac{1}{6!}$\Comment{Hard coded first two terms for $c_4$.}
	\State $c_5 \gets \frac{1}{5!} - z \cdot \frac{1}{7!}$
	\State $\bar z \gets -z$
	\State $p \gets \bar z$		\Comment{$p$ will the $(-z)^j$ factor in the loop.}
	\State $k \gets 8$		\Comment{Third term in $c_4$ contains factor $\frac{1}{8!}$.}

	\Repeat
		\State {$c_{4\rm, prev} \gets c_4$} \Comment{Keep old value to check for convergence.}

		\State $p \gets p \cdot \bar z$
		\State $c_4 \gets  c_4 + p \cdot \frac{1}{k!}$ \Comment{$1/k!$ comes from lookup table.}

		\State $k \gets k+1$
		\State $c_5 \gets  c_5 + p \cdot \frac{1}{k!}$
		\State $k \gets k+1$
	\Until {$c_4 = c_{4\rm, prev}$ }  \Comment{Converged?}
	\State $c_3 \gets \frac16 - z \cdot c_5$\Comment{Use Eq.~\ref{eq:recur} to get $c_3$, $c_2$ and $c_1$.}
	\State $c_2 \gets \frac12 - z \cdot c_4$
	\State $c_1 \gets 1 - z \cdot c_3$
	\While{$n>0$}			\Comment{Apply quarter angle formula $n$ times.}
		\State $z \gets 4 \cdot z$
		\State $c_5 \gets \frac1{16}\cdot (c_5+c_4+c_3+c_2) $
		\State $c_4 \gets \frac18\cdot  c_3 \cdot (1-c_1)$
		\State $c_3 \gets \frac16 -z \cdot c_5$
		\State $c_2 \gets \frac12 -z \cdot c_4$
		\State $c_1 \gets 1 -z \cdot c_3$
		\State $n\gets n-1$
	\EndWhile
	\State $c_0 \gets 1 -z \cdot c_2$
\end{algorithmic}\vspace{0.2cm}

%%%%%%%%%%%%%%%%%%%%%%%%%%%%%%%%%%%%%%%%%%%%%%%%%
\subsection{Implementation of Gau{\ss} $f$ and $g$-functions}\label{sec:gauss}
The precise implementation of Gauss $f$ and $g$ functions matters for long term integrations.
The straightforward implementation following \citep{MikkolaInnanen1999} leads to the $f$ and $g$-functions in Eq.~\ref{eq:fg}.
Note that for timesteps smaller than half an orbital period, the term $MG_2/r_0$ in $f$ is small compared to the first term (which is just 1). 
The same argument holds true for $\dot g$.
We can define new $\hat f$ and $\hat{\dot g}$-functions
\begin{eqnarray}
&\hat f = - M \frac{G_2}{r_0}\quad & \dot f = -\frac{M\,G_1}{r_0r} \label{eq:fghat}\\
&g = dt - M G_3            & \hat {\dot g} = -\frac{M\,G_2}{r}.
\end{eqnarray}
This allows us to rewrite the last step in solving the Kepler problem as
\begin{eqnarray}
\mathbf{r} = \left( \hat f \mathbf{r}_0 + g \mathbf{v}_0 \right) + \mathbf{r}_0 \quad\quad\quad\quad
\mathbf{v} = \left( \dot f \mathbf{r}_0 + \hat{\dot g} \mathbf{v}_0 \right) + \mathbf{v}_0.
\end{eqnarray}
Although this step is algebraically equivalent to the original Eq.~\ref{eq:fgupdate}, we achieve higher precision. 
The reason is that we can now ensure that the small quantities in brackets are summed before they are added to the larger quantity (the initial value).
We implement the same trick for the variational equations, Eqs.~\ref{eq:var1} and~\ref{eq:var2}

%%%%%%%%%%%%%%%%%%%%%%%%%%%%%%%%%%%%%%%%%%%%%%%%%
\subsection{A full integration in Jacobi coordinates} \label{sec:intinjac}
The algorithms to convert to and from Jacobi coordinates that we describe in Sect.~\ref{sec:jacobi} are unbiased and fast.
Nevertheless, we aim to avoid as many conversion as possible.

As it turns out, we can reduce the number of conversions per timestep to two, one for the positions from Jacobi coordinates to the inertial frame, and one for the accelerations from the inertial frame to Jacobi accelerations.
But note that this is only possible under the following assumptions:
1) the particle position and velocities are not changed in-between timesteps, e.g. manually by the user or by collisions,
2) outputs are not required at every timestep,
3) variational equations are not integrated,
4) no additional velocity-dependent forces are present.
In such a case, an integration starting from an arbitrary inertial frame is achieved as follows:\vspace{0.2cm}
\begin{algorithmic}
	\State calculate Jacobi coordinates
	\State drift all particles under $H_{\rm Kepler}$ for half a timestep, $\mathit{dt}/2$
	\While{$t<t_{\rm max}$}
		\State calculate 1st part of $H_{\rm Interaction}$ in Jacobi coordinates
		\State update positions in the inertial frame
		\State calculate 2nd part of $H_{\rm Interaction}$ in inertial frame
		\State convert accelerations from 2nd part to Jacobi accelerations
		\State apply kick from Jacobi accelerations to Jacobi velocities
		\If{not last timestep}
			\State drift all particles under $H_{\rm Kepler}$ for a full timestep $\mathit{dt}$
		\EndIf
	\EndWhile
	\State drift all particles under $H_{\rm Kepler}$ for half a timestep $\mathit{dt}/2$
	\State update both positions and velocities in the inertial frame.
\end{algorithmic}\vspace{0.2cm}
Note that we never update the velocities in the inertial frame until the end of the simulation (or when an output is needed).
We only convert the positions and velocities to Jacobi coordinates at the very beginning and not at every timestep.
Besides the obvious speed-up, avoiding to go back and fourth between different coordinate systems reduces the build-up of round-off errors and thus makes the integrator more robust.

%%%%%%%%%%%%%%%%%%%%%%%%%%%%%%%%%%%%%%%%%%%%%%%%%
\subsection{LCN calculation}\label{sec:megnocalc}
To calculate the Lyapunov characteristic number and the Lyapunov timescale we need to perform a linear least square fit to the function $Y(t)$.
Thus we need the mean and the covariance of $Y(t)$.
Storing all previous values of $Y(t)$ just to calculate its mean and covariance is inefficient.
We therefore implement an efficient one-pass method described by \cite{Pebay2008}.
This method lets us calculate the LCN at every timestep in $\mathcal{O}(1)$ and has the further advantage of being numerically more robust than the standard implementation.

\begin{figure*}
 \centering \resizebox{0.99\textwidth}{!}{\includegraphics{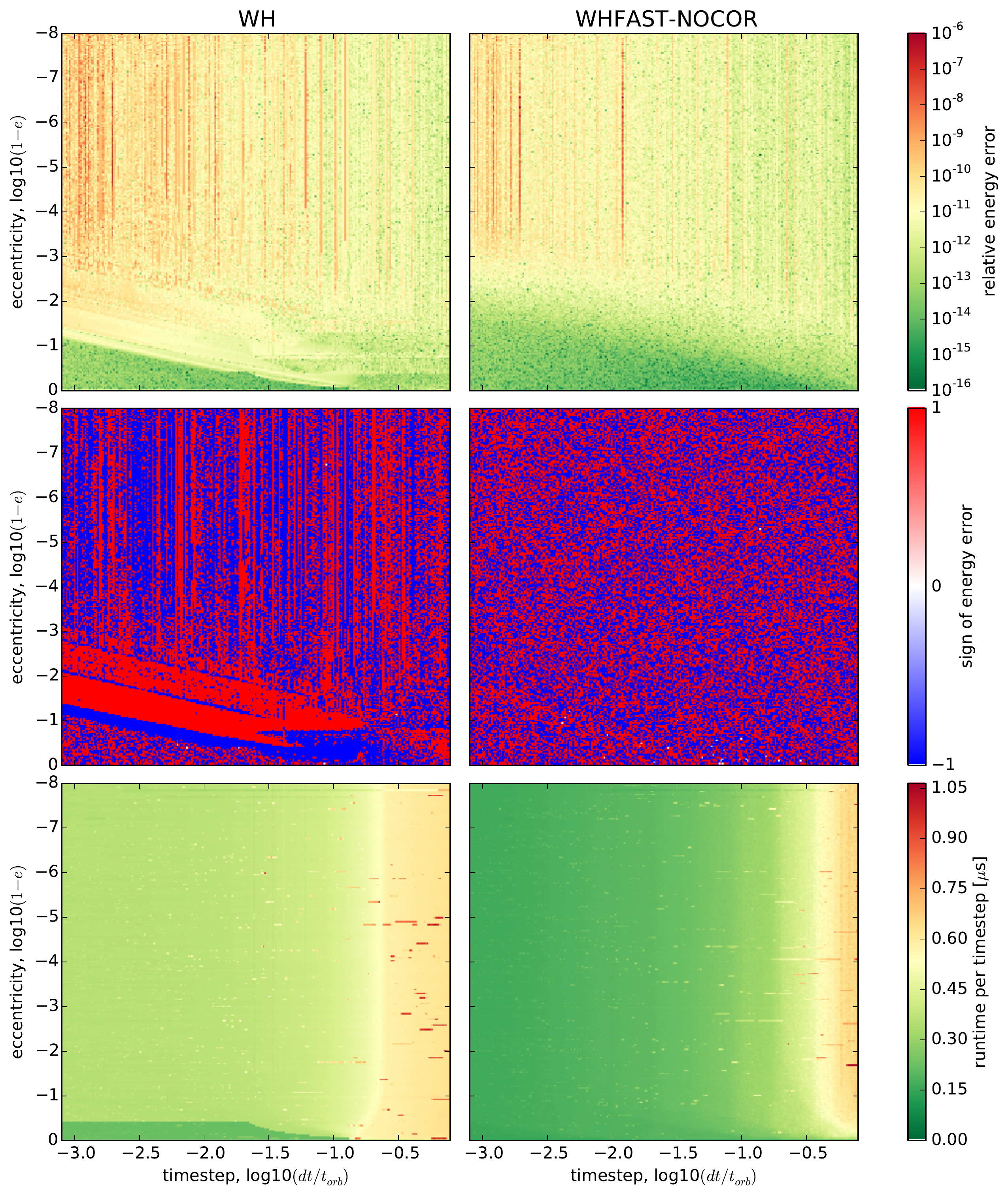}}
 \caption{
Tests of the Kepler-solver.
Simulation with two bodies, integrated for 100 orbits with varying eccentricity and timestep. Left column: results using the standard \wh integrator. Right column: results using our new \whf integrator. Top row: relative energy error at the end of the simulation.
Middle row: sign of the energy error at the end of the simulation.
Bottom row: average runtime for one timestep.
\label{fig:2body}}
\end{figure*}

%%%%%%%%%%%%%%%%%%%%%%%%%%%%%%%%%%%%%%%%%%%%%%%%%
%%%%%%%%%%%%%%%%%%%%%%%%%%%%%%%%%%%%%%%%%%%%%%%%%
%%%%%%%%%%%%%%%%%%%%%%%%%%%%%%%%%%%%%%%%%%%%%%%%%
\section{Numerical Results}\label{sec:numericalresults}
In this section, we test the speed, accuracy and numerical stability of \whf and compare it to other publicly available and widely used integrators. 
We begin by briefly defining our nomenclature for these other integrators and summarizing their properties.

\mer is a mixed-variable symplectic integrator implemented in fortran and provided by the \mer package \citep{Chambers1997}. 
This Wisdom-Holman style integrator uses high-order symplectic correctors. 
We directly call the fortran code without any modifications. 

\swifterwhm is again a classical 2nd-order Wisdom-Holman integrator without symplectic correctors \citep{WisdomHolman1991}. 
We use the integrator provided by the \swifter package.
It is implemented in fortran and we directly call the \swifter executable without any modifications.

\swifterhelio is also 2nd-order symplectic integrator without symplectic correctors \citep{Duncan1998}. 
It uses democratic heliocentric coordinates. 
We again use the integrator provided by the \swifter package.
It is implemented in fortran and we directly call the \swifter executable without any modifications.

\swiftertu is a 4th-order symplectic integrator. 
It is \emph{not} a Wisdom-Holman integrator but splits the Hamiltonian in kinetic and potential terms \citep{Gladman1991}. 
We also use the integrator provided by the \swifter package.
It is implemented in fortran and we directly call the \swifter executable without any modifications.

For a more direct comparison, we also make use of an integrator that we simply refer to as \wh. It is based on the \swifterwhm integrator in \swifter but ported to C and available in the \reb \citep{ReinLiu2012} package.
Like the \swifterwhm integrator, it is a symplectic integrator that works in the heliocentric frame, and does not implement any symplectic correctors. 
Note that this is not the original integrator used by \cite{WisdomHolman1991}, which is not publicly available.

\whf is C99 compliant. The C99 standard guarantees that floating point operations are not re-ordered by the compiler (unless one of the fast-math options is turned on).
Because of that, the final positions and velocities of particles agree down to the last bit across different platforms.
This makes \whf platform independent and the simulation results reproducible.
We verified this on different architectures (Linux, MacOSX), different CPUs (Intel Core i5-3427U, Intel Xeon E5-2697 v2, Intel Xeon E5-2620 v3) and different compilers (Apple LLVM 6.1.0, gcc 4.4.7).

%%%%%%%%%%%%%%%%%%%%%%%%%%%%%%%%%%%%%%%%%%%%%%%%%
\subsection{Two-body Kepler Solver}

The kernel of every Wisdom-Holman integrator is the Kepler solver. 
We describe our implementation in detail in Sections~\ref{sec:newton}-\ref{sec:gauss}.
Here, we test the Kepler solver using a two-body problem. 
The two body problem is invariant with respect to rescaling of the total mass, the mass ratio, the value of the gravitational constant and the orbital period.
What does matter is the eccentricity of the orbit and the ratio of the timestep to the orbital period. 
We thus scan the parameter space in those two dimensions by integrating two bodies for 100 orbital periods.
We explore an extremely wide parameter space.
The eccentricities range from zero to $0.999\,999\,99 = 1-10^{-8}$.
The range of timesteps goes from 0.1\% of the orbital period all the way up to one orbital period.

Fig.~\ref{fig:2body} shows the performance of \whf (right column) compared to \wh (left column).
The top row shows the absolute value of the relative energy error at the end of the simulation.
The middle row shows the sign of the energy error.
The bottom row shows the average runtime for a single timestep.
The vertical lines visible in the top row correspond to timestep resonances \citep{WisdomHolman1992,ToumaWisdom1993,Rauch1999}.

One can see that \whf is significantly more accurate than the standard \wh integrator for the most important parts of parameter space (eccentricities less than $\sim 0.99$).
The relative energy is conserved better by two to three orders of magnitude.
Most importantly, note that the energy error in the standard \wh integrator is biased over large regions of the parameter space (there are large blue and red areas in the second row).
On the other hand, \whf has a random energy error throughout the parameter space.
Having a biased energy error will lead to a long-term linear growth of the energy error (see below).

In the entire parameter space explored, \whf requires less time to complete a timestep than \wh.
The speed-up is typically between 20\% and 100\%.
For the integrations performed in this section, we convert to and from Jacobi coordinates at every timestep to provide a fair comparison. 
Thus, the speed-up and the energy-conservation properties of \whf are in fact even better than shown here in any actual production run (see Sect.~\ref{sec:intinjac}).

%%%%%%%%%%%%%%%%%%%%%%%%%%%%%%%%%%%%%%%%%%%%%%%%%
\subsection{Short Term Energy Conservation}\label{sec:shorttermenergy}
\begin{figure}
 \centering \resizebox{0.99\columnwidth}{!}{\includegraphics{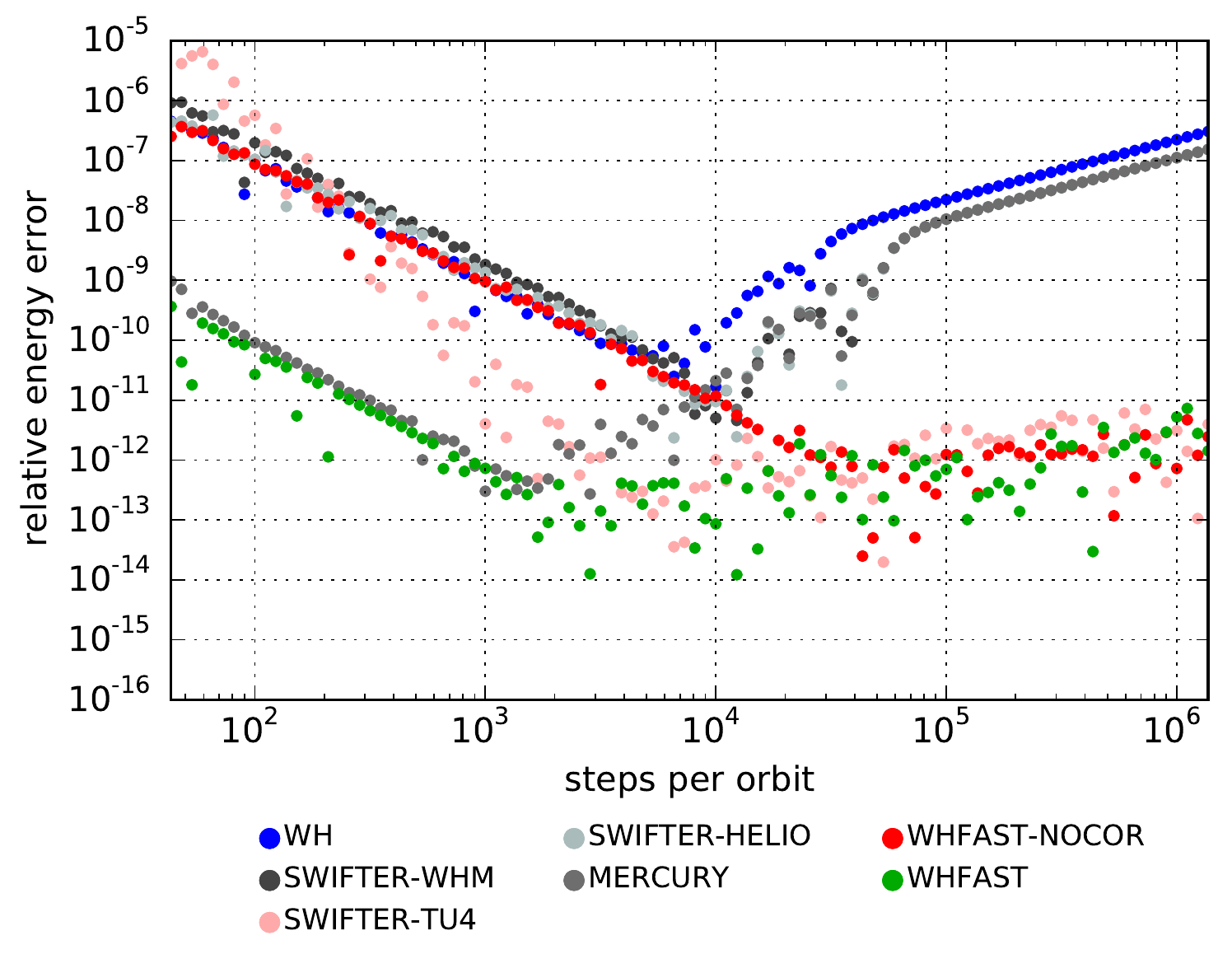}}
 \caption{
Relative energy error in simulations of the outer Solar System after 1000~Jupiter orbits as a function of the number of steps per orbit.
\label{fig:shorttermenergy}}
\end{figure}

To compare the accuracy of the different integrators in a realistic test case, we run simulations of the outer Solar System for one thousand Jupiter orbits ($12\,000$~years).
We include the Sun and four massive bodies with approximate initial conditions corresponding to those of Jupiter, Saturn, Uranus and Neptune. 
In each simulation the initial conditions and masses are randomly perturbed by 0.1\%.
In Fig.~\ref{fig:shorttermenergy}, we plot the relative energy errors at the end of the simulation as a function of the number of timesteps imposed per Jupiter orbit.

One can see that all the integrators except \swiftertu are second-order schemes. 
For timesteps between 20\% and 0.1\% of the orbital period of Jupiter (50 to 1000~timesteps per orbit), their error decreases quadratically with decreasing timesteps. 
This is the error term $E_{\rm bound}$ introduced in Sect.~\ref{sec:errors}.

However, decreasing the timestep also increases the number of floating point operations.  There will therefore be a timestep value at which the numerical round-off error dominates over the error associated with the symplectic method itself $E_{\rm bound}$. 
For that reason we find that for small timesteps, less than 0.1\% of the shortest orbital period, the errors of all integrators rise instead of decreasing further.
Thus there is an optimum timestep $\mathit{dt}_{\rm opt}$ that yields the minimum energy error.  
This optimum timestep depends on the length of the integration and will be larger for longer simulations.

In Fig.~\ref{fig:shorttermenergy} one can see that the errors of \wh, \swifterwhm, \swifterhelio and \mer rise very rapidly after reaching $\mathit{dt}_{\rm opt}$, scaling as at least $\mathit{dt}^{-2}$ for the first decade. 

The optimum timestep for \whf is roughly 0.1\% of the shortest orbital period.
However, \whf's error grows much more slowly with decreasing timestep than that of the other second-order integrators. 
In fact, the error is dominated by $E_{\rm rand}$ and thus follows $\mathit{dt}^{-1/2}$ as the number of timesteps $N_{\rm steps}$ increases as $\sim\mathit{dt}^{-1}$ if we keep the total integration time constant.
Thus the behaviour of \whf in Fig.~\ref{fig:shorttermenergy} for small timesteps can be seen as the first indication that \whf follows Brouwer's law (see Sects.~\ref{sec:errors} and~\ref{sec:longtermtest}). 

The \swiftertu integrator is the only other integrator we tested that seems to follow Brouwer's law, but it performs poorly at large timesteps. 
This is expected, since unlike the other integrators, \swiftertu does not assume a Keplerian splitting and must therefore take smaller timesteps to accurately reproduce the orbital motions.

Integrators with symplectic correctors, \mer and \whf, perform significantly better for long timesteps.
Their energy conservation is three orders of magnitude better ($E_{\rm bound}$ is three orders of magnitude smaller) compared to integrators without symplectic correctors. 
This is due to the mass ratio of Jupiter and the Sun being roughly $10^{-3}$.
The order of the symplectic corrector is not very important for relatively high mass ratios such as these, i.e. a fifth-order symplectic corrector performs as well as an 11th-order one. 
For much smaller mass ratios (when the mass ratio is less than the timestep ratio), higher-order symplectic correctors are advantageous.

Note that $\mathit{dt}_{\rm opt}$ for almost all of the integrators is $10^{-3}$~orbital periods of Jupiter, i.e. 4 days.
This is significant because Mercury's orbital period is 88~days.
Thus if we included Mercury in our simulation, we would be very restricted in our timestep choice.
We need more than 20 timesteps ($\mathit{dt}\approx4$~days) to resolve Mercury's orbit accurately. 
However, if we choose choose a timestep smaller than 4 days, we start to accumulate errors in the outer Solar System. 
It is worth reiterating that the simulations shown in Fig.~\ref{fig:shorttermenergy} all ran for only 1000~orbits.
If we ran a longer simulation with the same timestep, we would have more timesteps and thus accumulate more round-off errors by the end of the simulation.
One can therefore reach better energy conservation with a longer timestep.
In other words, $\mathit{dt}_{\rm opt}$ is larger for longer integration times.

%%%%%%%%%%%%%%%%%%%%%%%%%%%%%%%%%%%%%%%%%%%%%%%%%
\subsection{Speed Comparison}\label{sec:speedcomparison}
\begin{figure}
 \centering \resizebox{0.99\columnwidth}{!}{\includegraphics{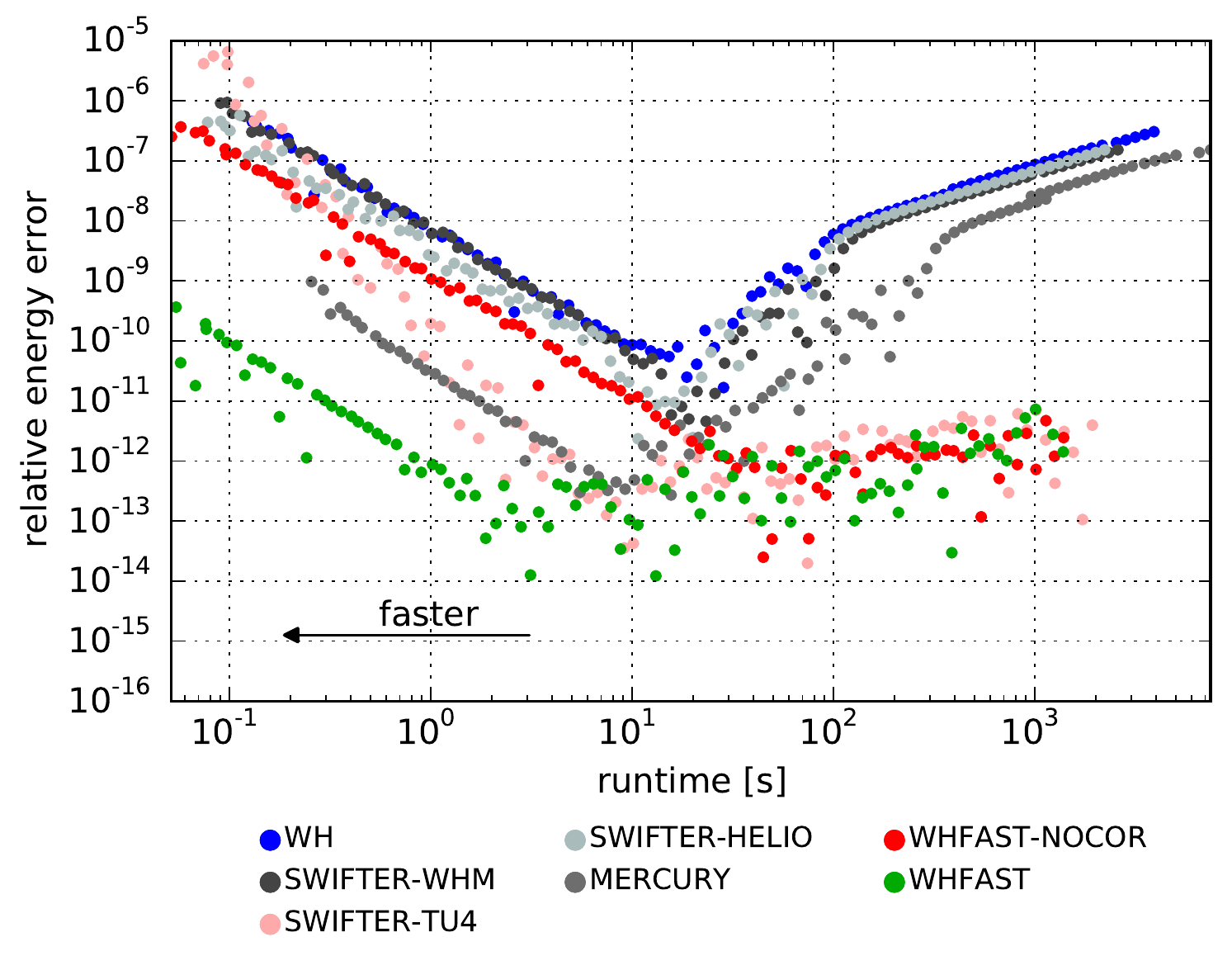}}
 \caption{
Relative energy error in simulations of the outer Solar System after 1000~Jupiter orbits as a function of run time.
\label{fig:speedcomparison}}
\end{figure}

We run the same simulations as in Sect.~\ref{sec:shorttermenergy} to compare the speed of the different integrators.
Fig.~\ref{fig:speedcomparison} shows the relative energy error as a function of runtime.
The results show that no matter what the desired energy error is, \whf is the fastest integrator.
In the large timestep limit, the speed-up compared to \mer is roughly a factor of~5.

In the small timestep limit, $\mathit{dt} < \mathit{dt}_{\rm opt}$, we can only compare \whf to \swiftertu, as all the other integrators' errors are significantly larger (by 4 to 5~orders of magnitude) due to numerical roundoff errors (see below).
\swiftertu is as fast for small timesteps as \whf but, as noted above, is unsuitable for large timesteps since it is not a Wisdom-Holman integrator.
It is only shown here as a comparison.

%%%%%%%%%%%%%%%%%%%%%%%%%%%%%%%%%%%%%%%%%%%%%%%%%
\subsection{Long Term Energy Conservation}\label{sec:longtermtest}
\begin{figure*}
 \centering \resizebox{0.99\textwidth}{!}{\includegraphics{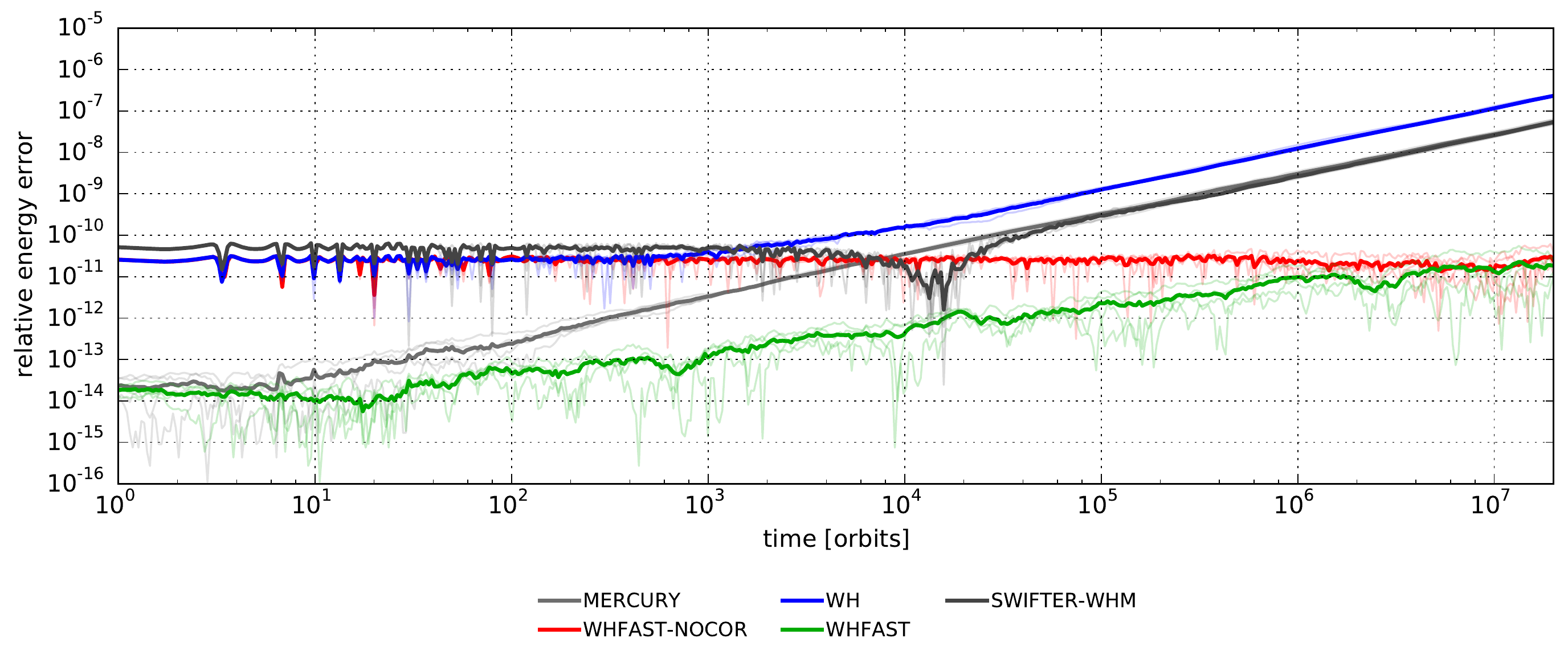}}
 \caption{
Relative energy error in simulations of the outer Solar System as a function of time for different symplectic integrators. The timestep for all simulations is 1.5~days. 
\label{fig:longtermtest}}
\end{figure*}

Let us finally address the most important benchmark, the long term energy conservation properties of \whf compared to other integrators in a real world test case.
In this section we only study the energy error, but other conserved properties like the angular momentum behave the same way.
In Fig.~\ref{fig:longtermtest}, we show the time evolution of the relative energy error in a simulation of the outer Solar System.
As in Sect.~\ref{sec:shorttermenergy}, we include the Sun and four massive bodies with approximate initial conditions corresponding to those of Jupiter, Saturn, Uranus and Neptune. 
The timestep for all simulations is 1.5~days.
Note that this timestep is smaller than what one would typically choose for this kind of integration.
However, with a 1.5~day timestep we reach machine precision for integrators that use symplectic correctors, allowing us to better quantify the long-term behaviour of \whf.
All the effects we discuss here are also present in simulations with longer timesteps, but they would manifest themselves in the relative energy error at a later time.
We run four simulations for each integrator and randomly perturb the initial conditions and masses by 0.1\% in each simulation. 
We plot the individual simulations as thin lines, and the average error as a bold line.

The lower relative energy bound is set by machine precision for all integrators, roughly $10^{-16}$.
\whf and \mer, the integrators in our sample that have symplectic correctors, almost reach this limit early on in the simulation.
The bound energy error $E_{\rm bound}$ is approximately $10^{-14}$.
The integrators without symplectic correctors, \swifterwhm and \wh have an energy error roughly three order of magnitudes higher $E_{\rm bound}\approx 10^{-10.5}$.

From Fig.~\ref{fig:longtermtest} it is clear that the integrators \mer, \wh and \swifterwhm show a linear behaviour in the energy error at late times.
This is due to the term $E_{\rm bias}$.
The $E_{\rm bias}$ term already dominates at early times (after 100 Jupiter orbits) for \mer because the symplectic correctors lower the value of $E_{\rm bound}$.
For \wh and \swifterwhm the $E_{\rm bias}$ term dominates after $10\,000$~Jupiter orbits.
This result shows that one or more steps in these integration algorithms are biased.
We found that the two main contributions were the inaccurate implementation of the rootfinder for Kepler's equation and the conversions to and from Jacobi coordinates.
In \whf, $E_{\rm bias}$ is absent, showing that its implementation is completely unbiased.

Since all integrators are implemented in double floating-point precision and use the same timestep, they all have roughly the same error term  $E_{\rm rand}$.
However, it is only visible in Fig.~\ref{fig:longtermtest} for the \whf integrator.
For all other integrators the linearly growing term $E_{\rm bias}$ dominates over $E_{\rm rand}$.
 
If we increase the timestep, the linear error growth will show up at a later time because $E_{\rm bound}$ will be larger.
However, it is still present at all times.
Let us think of a symplectic integrator as an exact integrator for a perturbed Hamiltonian $\tilde{\mathcal{H}}$ with high frequency terms added compared to $\mathcal{H}$ in Eq.~\ref{eq:H}, see e.g. \cite{Wisdom1996}.
Then the quantity related to the energy error for $\tilde{\mathcal{H}}$, let us call this $\tilde{E}$, should be conserved exactly at all times (that is the idea of a symplectic integrator).
However, if the implementation is biased, $\tilde{E}$ will undergo a linear growth at all times.
With \whf, we improve the conservation of $\tilde{E}$ by many orders of magnitude in any integration, regardless of timestep.

This difference could have important implication for the dynamical evolution of the system and could for example push it from a stable to an unstable region of parameter space. 
We plan to study the effect of different integrators on systems near a chaotic/non-chaotic separatrix in a follow up paper.

%%%%%%%%%%%%%%%%%%%%%%%%%%%%%%%%%%%%%%%%%%%%%%%%%
%%%%%%%%%%%%%%%%%%%%%%%%%%%%%%%%%%%%%%%%%%%%%%%%%
%%%%%%%%%%%%%%%%%%%%%%%%%%%%%%%%%%%%%%%%%%%%%%%%%
%%%%%%%%%%%%%%%%%%%%%%%%%%%%%%%%%%%%%%%%%%%%%%%%%
\section{Conclusions}\label{sec:conclusions}
In this paper, we presented \whf, a new implementation of a symplectic Wisdom-Holman integrator.
Key advantages and improvements over other publicly available implementations of symplectic integrators are:

\textit{\whf is faster by a factor of $1.5$~to~$5$.} 
Of that, a 50\% speedup comes from the improved Kepler solver, where we use a fast convergence criteria for Newton's method and an efficient implementation of $c$ and $G$-functions.
The remainder of the speedup is due to combining drift steps at the end and beginning of each timestep and to only converting to and from Jacobi coordinates when needed.

\textit{The Kepler solver is more accurate and unbiased.}
We achieve this thanks to improvements to the convergence criteria in Newton's method, a Laguerre-Conway solver for highly eccentric orbits with long timesteps, the high accuracy implementations of the $c$ and $G$-functions and a careful ordering of floating-point operations.

\textit{We remove the secular energy error that grows linearly with integration time.}
This is due to two improvements. 
First, the unbiased Kepler solver.
Second, the improved and also unbiased coordinate transformations to and from Jacobi coordinates.
To our knowledge, \whf is the first publicly available implementation of a Wisdom-Holman integrator that follows Brouwer's law over long timescales for small enough timesteps and does not show a linear growth in the energy error.

\textit{We implement variational equations that allow us to compute the Lyapunov timescale and the MEGNO.}
Our algorithm to calculate the Lyapunov timescale uses a numerically stable algorithm that is based on a one-pass covariance filter.
The variational equations do not require us to solve Kepler's equation and are thus very inexpensive to calculate.

\textit{Symplectic correctors of order 3, 5, 7, and 11 are implemented.}
These symplectic corrector allow for high-accuracy simulations of systems with small mass ratios.
Even for relatively massive planets like those in the Solar System, symplectic correctors achieve an improvement of three orders of magnitude.
For long integrations, the performance cost of symplectic correctors is negligible and so our default setting uses an 11th-order corrector.

\textit{\whf lets the centre-of-mass move freely during an integration.}
We integrate an additional degree of freedom in order for our integrator to work in any inertial frame, i.e. one is not restricted to the heliocentric or barycentric frame. 
Additionally, we do not tie our implementation to a specific choice of units.

\textit{The integrator is available as an easy to use python module.}
The module works on both python 2 and 3.
It can be installed on most Unix and MacOS systems with a single command:
\begin{lstlisting}
pip install rebound
\end{lstlisting}
\noindent The following python script imports the rebound module, adds particles to the simulation, selects an integrator and timestep and runs the integration.
\begin{lstlisting}
import rebound
rebound.add(m=1)
rebound.add(m=0.001, a=1.)
rebound.add(m=0.001, a=2., e=0.1)
rebound.integrator = 'whfast'
rebound.dt = 0.01
rebound.integrate(6.2831)
\end{lstlisting}
More complicated examples and the source code of \whf (written in C, compliant with the C99 standard) can be found in the \reb package.
\reb includes several other integrators, collision detection algorithms, a gravity tree code and much more.
The \reb git repository is hosted at \url{https://github.com/hannorein/rebound}.

We also provide an experimental hybrid integrator for simulations in which close encounters occur. 
The hybrid integrator switches over to a high-order non-symplectic integrator \citep[IAS15,][]{ReinSpiegel2015} during a close encounter.
A detailed discussion of this integrator and its properties will be given in a follow-up paper.

We hope that with the speed and accuracy improvements, \whf will become the go-to integrator package for short and long-term orbit simulations of planetary systems.

%%%%%%%%%%%%%%%%%%%%%%%%%%%%%%%%%%%%%%%%%%%%%%%%%
\section*{Acknowledgments}
This research has been supported by the NSERC Discovery Grant RGPIN-2014-04553.
We thank Wayne Enright, Philip Sharp and Scott Tremaine for stimulating discussions and Jack Wisdom for a helpful referee report.

\bibliography{full}
\end{document}